\documentclass[a4paper,11pt]{article}

\usepackage[english]{babel}
\usepackage[utf8]{inputenc}
\usepackage{amsmath}
\usepackage{graphicx}
\usepackage[colorinlistoftodos]{todonotes}
\usepackage{siunitx}

\usepackage{graphicx}
\usepackage{booktabs}
\usepackage{caption}
\usepackage{subcaption}
\usepackage{url}
\usepackage{algorithm}
\usepackage{algpseudocode}
\usepackage{tabularx}
\usepackage{float}
\usepackage[colorinlistoftodos]{todonotes}
\usepackage[colorlinks=true, allcolors=blue]{hyperref}
\usepackage{authblk}
\usepackage[round]{natbib}

\author[1]{Mehmet Tan \thanks{Corresponding author:mtan@etu.edu.tr}}
\author[1]{Ozan Fırat Özgül}
\author[1]{Batuhan Bardak}
\author[1]{Işıksu Ekşioğlu}
\author[2]{Suna Sabuncuoğlu}
\affil[1]{Department of Computer Engineering, TOBB University of Economics and Technology, Ankara, Turkey}
\affil[2]{Toxicology Department, Faculty of Pharmacy, Hacettepe University, Ankara, Turkey}
\date{}

\title{Drug response prediction by ensemble learning and drug-induced gene expression signatures\footnote{This research was supported by The Scientific and Technological Research Council of Turkey (Grant no:115E274)}}

\date{}

\begin{document}
\maketitle

\begin{abstract}

Chemotherapeutic response of cancer cells to a given compound is one of the most fundamental information one requires to design anti-cancer drugs.  Recently, considerable amount of drug-induced gene expression data has become publicly available, in addition to cytotoxicity databases. These large sets of data provided an opportunity to apply machine learning methods to predict drug activity. However, due to the complexity of cancer drug mechanisms, none of the existing methods is perfect. In this paper, we propose a novel ensemble learning method to predict drug response. In addition, we attempt to use the drug screen data together with two novel signatures produced from the drug-induced gene expression profiles of cancer cell lines. Finally, we evaluate predictions by \emph{in vitro} experiments in addition to the tests on data sets. The predictions of the methods, the signatures and the software are available from \url{http://mtan.etu.edu.tr/drug-response-prediction/}.




\end{abstract}

\section{Introduction}

The highly variable nature of treatment responses among cancer patients necessitates tailor-made therapies. 
Forecasting the most feasible drug therapy for an individual patient, being the vital part of such personalized 
therapies, is a gruelling task since it requires the application of several candidate drugs on cancer on an individual basis.
Determining chemotherapeutic response of malign cells \emph{in silico} is an important problem whose solution  
can greatly contribute to both drug discovery and to such personalized treatment design. Several recently introduced 
large scale drug screen databases provided a significant opportunity to train machine learning models 
that predict drug activity \citep{garnett2012,Barretina2012TheSensitivity}. 
These databases provide data based on a large number of cytotoxicity experiments 
treating cancer cell lines with chemicals. Accurate machine learning models exploiting these databases 
have the potential to replace some of the wet-lab experiments performed for these different procedures. 

There are several recent studies on drug sensitivity and anti-cancer drug response prediction. In \citep{Wan2014AnChallenge}, 
random forest regression model is used for obtaining a single drug or a group of drugs for a certain cancer. 
\citep{Zhang2015PredictingModel} proposes an approach with cell line similarity network (CSN) and drug similarity 
network (DSN) data to predict anti-cancer drug responses of a given cell line by using dual-layer integrated cell 
line-drug network. To predict drug activity, \citep{Turki2017APrediction} proposes a novel approach that uses link 
prediction algorithms and compare their results with existing drug sensitivity prediction models. Based on their 
experimental results, link prediction algorithms perform better than the methods used in comparison. Results based on a 
similar motivation is reported in \citep{Stanfield2017DrugProblem} where the authors compute network profiles derived from 
a heterogeneous drug response network. They, then train a classifier to distinguish sensitive and resistant cell lines. In a recent 
study \citep{Zhang2018ALines}, the authors integrate information from gene expression profiles, 
drug chemical structures, drug-target data and protein protein interactions to derive a heterogeneous network for
predicting cell line-drug associations, where the novel results are validated by the literature evidence.
Multi-task learning schemes have come into prominence as a consequence of their capability of exploiting inter-drug 
relationships during training. While multi-task learning for the drug activity prediction was largely limited to regularized 
linear models before, recently more sophisticated approaches have emerged. 
An approach called Kernelized Bayesian Multi-task Learning (KBMTL) \citep{Gonen2014} relying on kernel-based dimension 
reduction and multi-task learning demonstrated notable performance for drug response prediction.
The study by Tan \citep{Tan2016PredictionLearning} has shown that the trace-norm regularized multi-task learning 
followed by a kernel transform of the input gene expression data is superior to several other methods proposed 
for this task. A similar work by Yuan et al \citep{Yuan2016MultitaskSensitivity}, also demonstrates the improved 
predictive power of multi-task learning with 
trace-norm regularization. In another study \citep{Wang2017ImprovedRegularization}, the authors premediate 
that the similar drugs should possess similar responses and define drug-similarity by Pearson correlation 
of drug responses. This is followed by application of a novel method called similarity-regularized matrix factorization (SRMF) 
which maps drugs and cell lines to a lower dimensional space and reconstruct the drug response matrix in this latent space.
The usage of ensemble methods were also found promising in drug response prediction task, as in many other problems. 
These low-variance models surpassed all other schemes in a recent DREAM challenge on predicting human responses to 
toxic compounds \citep{Eduati2015PredictionCompetition}. Winners of both sub-tasks utilized ensemble of random forest 
models where each forest was devoted to a single cluster of cell lines. In addition to ensemble learning, 
biologically-plausible preprocessing techniques such as filtrating irrelevant genetic features, clustering cell 
lines and chemical using external data sets were found to be efficient strategies. None of these methods
consider to use both multi-task and single-task methods together in an ensemble model. Also none 
considered to integrate the LINCS database to the transcriptional profiles for this task. 

Our contributions in this paper can be summarized as follows. First, we propose to use an ensemble 
model to predict cytotoxicity of anti-cancer drugs by using four different prediction methods. 
Three of these methods have been recently proposed for this task, and the last one is a multi-task 
neural network. For each of these, similar cell lines and drugs are grouped and a large number 
of models are trained on these subsets of data as in the winning strategy for DREAM challenge 
\citep{Eduati2015PredictionCompetition}. We, then, combine the predictions from these models 
into a single prediction by stacked generalization. Second, we propose cell line sensitivity and 
drug activity signatures by integrating multiple databases and study the effect of 
these to the response prediction performance. Finally, we validate the predictions \emph{in vitro} 
by treating one cell line with five selected compounds. The results are promising; both the ensemble 
model and signatures improve the prediction performance compared to the baseline methods. 

In the following sections, we first describe the data sets that we use. Then, Section \ref{signatures-sec}
discusses the drug activity and cell line sensitivity signatures and Section \ref{ensemble-sec}
provides the details on the proposed ensemble method. Finally, we report the experimental results in 
Section \ref{experimental-sec} and conclude in Section \ref{conc-sec}.

\section{Data sets}

The first data set that we use is the Genomics of Drug Sensitivity in Cancer (GDSC) database
\citep{Iorio2016ACancer}. To produce GDSC database, more than a thousand cancer cell lines were screened against a 
large number of anti-cancer therapeutics. The main aim of the source is to identify biomarkers that 
lead to drug sensitivity or resistance. The cell lines are characterized in multiple omics features
such as the gene expression, copy number variation and coding variants. We utilize the gene expression data
as there is a wide consensus that this is the most informative data type for drug response prediction
\citep{Costello2014}. Gene expression data is produced by Affymetrix Human Genome U219 Array and normalized 
by robust multiarray averaging (RMA). The second data set that we utilize is the Cancer Cell Line Encyclopedia (CCLE) 
\citep{Barretina2012TheSensitivity}.
This is a similar screen to GDSC where the mRNA expression data is produced by using AffyMetrix 
U133+2 arrays. Provided data is the perturbational profile of 504 cancer cell lines treated with 24 
anti-cancer drugs. Gene expression profiles provided by GDSC and CCLE is high-dimensional; the number of genes 
is around 17000. To reduce dimension, we applied the gene selection procedure we previously proposed in 
\citep{Tan2016PredictionLearning}, where the number of genes selected is 1274 and 1239 for CCLE and GDSC respectively.
The last database we use, Library of Integrated Network-based Cellular Signatures (LINCS), is composed of drug-induced 
gene expression profiles of the cancer cell lines \citep{Subramanian2017AProfiles}. 
The database has recently been enlarged and moved 
to a new home called the CMAP L1000 database \footnote{http://clue.io}. The database currently 
holds perturbational profiles of tens of cell lines treated with tens of thousands of perturbagens
most of which are small molecules. The profiles are given in terms of 978 landmark genes that can explain 
most of the variance in the gene expression profile 
of a cell line. In this paper we used the LINCS Phase II version by using the provided API. 

\section{Signatures}
\label{signatures-sec}

This section describes the proposed signatures to represent the activity and sensitivity of 
the drugs and cell lines, respectively. 

\subsection{Drug activity signatures}

Lincscloud API provides access to the experimental results in Library of Integrated Network-based 
Cellular Signatures (LINCS)\footnote{http://www.lincsproject.org/} Program database. This database 
exhibits the effect of a drug on numerous cell lines by means of up or down regulations of landmark
genes chosen by the database curators. We use these experimental results with the aim of generating
drug activity signatures which represents each drug by its induced gene expression changes. 

The first step of drug activity signature generation is the formation of experimental signatures, 
$ExpSig(dr,CL)$, that represents the gene-specific changes on a cell line $CL$ upon application of 
drug $dr$ in vector form:

\vspace{-0.5cm}

\begin{equation}
ExpSig(dr,CL) = \langle gene_{1}\uparrow,gene_{2}\downarrow , gene_{3}\downarrow, ... , gene_{K}\uparrow  \rangle
\end{equation}

\noindent where the arrows represent the up and down regulations for the genes. 

For a given drug $dr$ and cell line $CL$, $ExpSig(dr,CL)$ can be constructed by a single LINCS 
access, where the database API provides the most regulated 50 among the 978 landmark genes.

Experimental signatures are generated for all cell line-drug pairs in our data set. In the LINCS database, multiple experiments can exist for a given cell line-drug pair with different experimental parameters such as dose and duration.  For instance, there are 188 different experiments for cell line MCF7 treated with Vorinostat, where we considered these as distinct experiments. 

Based on the experiment signatures ($ExpSig$), drug activity signature ($ActSig$) that represents the cumulative gene expression changes induced by a drug $dr$ is defined as follows:

\vspace{-0.3cm}

\begin{equation}
ActSig(dr) = \bigcup_{CL}ExpSig(dr,CL)
\end{equation}

\noindent where, $\bigcup_{CL}$ represents the combination of the experiment signatures for $dr$ over 
all experiments involving $dr$ in LINCS. Here, combination can be performed in several different ways. We define combination as the number of experiments for which the genes are up and down regulated. Therefore, 

\vspace{-0.3cm}

\begin{equation}
ActSig(dr) = \langle g_{1}(\uparrow n^u_1,\downarrow n^d_1), ... , g_{K}(\uparrow n^u _K,\downarrow n^d_K)  \rangle \\
\end{equation}

The notation $g_{1}(\uparrow n^u_1,\downarrow n^d_1)$, states that $dr$ is observed to up-regulate gene $g_1$ in $n^u_1$ experiments and it down-regulates $g_1$ in $n^d_1$
experiments.

$ActSig$ defined here, therefore, represents how a drug effects the cell lines, where this can be considered 
as an expression of mode of action (MOA) for a drug, in terms of gene regulations. $ActSig$ also has 
the property of decreasing the effect of experimental errors by averaging over several experiments.

\subsection{Cell line sensitivity signatures}

While GDSC and CCLE databases include the cytotoxicity measurements of compounds on particular cell lines, LINCS database exhibits differential gene expression profiles of cell lines upon treatment with a large number of compounds. By a cell line sensitivity 
signature (CLSS) we aim to gather the gene expression changes in a cell 
line $CL$, induced by different compounds. However, we restrict these compounds to be the ones that are found to be active against $CL$, where the activity information is extracted from other pharmacological screening studies such as GDSC. By this restriction, our purpose is 
to build a signature that describes drug-induced differential gene expression in the cell lines which are sensitive to those drugs (hence the name 'sensitivity signature'). 

Building the CLSS for a cell line is initiated by selecting an activity threshold, $T$, for the considered activity measure ($IC_{50}$ in our case). Then, for all the ($dr$,$CL$) pairs with $IC_{50}$ below $T$, we query the LINCS database to retrieve the up or down regulated landmark genes. Querying LINCS for ($dr$, $CL$) pair leads to three different possible scenarios: (i) LINCS might include ($dr$,$CL$) experiment and experimental results can be directly used, (ii) ($dr$,$CL$) experiment might not have been performed in LINCS, despite the existence of $dr$ and/or $CL$. In this case, the experimental results of $dr$ applied on cell lines similar to $CL$ can be used, (iii) $dr$ may not be available in LINCS. In this last case, if the (known) target proteins of $dr$ have been used in gene knock-out experiments in LINCS, their results on cell lines similar to $CL$ can be used to approximate the drug effect. Formally, the result of this 
procedure that we call the sensitivity signature (SensSig) can be defined as follows.

\begin{equation}
SensSig(dr,CL) = 
  \begin{cases}
      ExpSig(dr,CL) & \text{($dr,CL$) $\in$ LINCS} \\
      ActSig(dr) & \text{$dr$ $\in$ LINCS and $(dr,CL)$ $\notin$ LINCS} \\ 
      & \text{(on similar cell lines to $CL$)} \\
      TargetSig(Targets_{dr}) & \text{$dr$ $\notin$ LINCS (on similar cell lines to $CL$)} 
  \end{cases}
\end{equation}

\noindent where we used $\in$ to denote inclusion in database for ease of notation and 
$TargetSig(Targets_{dr})$ corresponds to case (iii) above, differential gene expression 
profile for the gene knock-out experiments for the targets of $dr$, $Targets_{dr}$.

As mentioned above, CLSS aims to cover the differential gene expression profile 
leading to cell death in tumor cells. $SensSig(dr,CL)$ describes the profile 
in response to $dr$ only. As there can be a large number of experiments on $CL$ 
in LINCS, combination of $SensSig(dr,CL)$ over all drugs applied on $CL$ can describe
the cumulative gene expression changes that are critical for $CL$. Based on this, we define 
CLSS as follows, 

\begin{equation}
CLSS(CL) = \bigcup_{dr} SensSig(dr,CL)
\label{clss-eq}
\end{equation}

The combination $\bigcup_{dr}$ in the above equation can be defined in a number of
different ways. This can simply be a set union. However, the one which we think can 
better represent the cumulative sensitivity, is to construct CLSS as a set of vectors,
where each vector corresponds to the union of signatures over drugs targeting the same pathway. 
The target pathway information for each compound is obtained from GDSC.
This way we will have a number of signatures for each cell line corresponding to the 
target pathways of the drugs in Equation \ref{clss-eq}. 


\subsection{Signature similarities}

As mentioned, the two signatures defined above characterize the cell lines and drugs in terms 
of sensitivity and activity respectively.  The sensitivity signature is a representation
of the gene expression changes that lead to cell death in a certain cell line. Similarly, drug activity 
signature is a representation of the "mode of action" for a drug, summarizing the cumulative differential 
gene expression upon treatment on cell lines. Therefore, a natural next step is to compare these 
signatures to predict whether a given drug is active on a given cell line. 

This approach is similar to the connectivity score of the Connectivity Map (CMAP) \citep{Lamb2006TheUsing} 
(the previous version of LINCS) for drug repositioning. The connectivity 
score defined in CMAP also compares two differential gene expression signatures to find a drug that 
has the effect of reverting the expression profile back to normal.  CMAP finds and outputs the drugs
whose effect is the opposite of a given differential gene expression 
profile for a certain condition (a disease for example). Our ultimate goal here, on the other hand, 
is to directly predict the drug activity. To this aim, we propose to use the similarity between 
the proposed drug and cell line signatures to be able to "help" some other methods to better 
predict drug response. In other words, we are looking for the drugs that will kill most of the malignant 
cells in a certain tumor. 

To sum up, we hypothesize that the matching between $CLSS(CL)$ and $ActSig(dr)$ should give us a clue 
on the activity of the drug $dr$ on the cell line $CL$. For this, we used cosine similarity between 
the signatures, where, among other choices, we observed that it performed better. As each $CLSS$ is 
a list of vectors (corresponding to different pathways), we computed the similarity of each 
vector in $CLSS(CL)$ with $ActSig(dr)$ and used the maximum of these as the similarity between the 
sensitivity of $CL$ and the activity of $dr$. So from here on, we will refer this maximum value 
as "signature similarity". 

\section{Ensemble model for drug activity prediction}
\label{ensemble-sec}

There can be multiple motivations to build ensembles as listed in \citep{Dietterich2000EnsembleLearning}; statistical,
representational and computational. All these different aspects more or less contribute to the motivation
of the work in this paper. Despite recent improvements in availability of drug response data, 
amount of data can still cause multiple different hypotheses to perform reasonably well, and combining them can 
improve the performance. Also, we do not yet know the best representation for drug response models which can differ 
for each individual drug, therefore an ensemble with different representation assumptions, again, might help. 
Computational aspect seems to contribute less than the formers to the motivation, however as the available 
data increase, we think it can become equally important as well. 

\begin{figure}[h]
\centering\fbox{\includegraphics[scale=0.8]{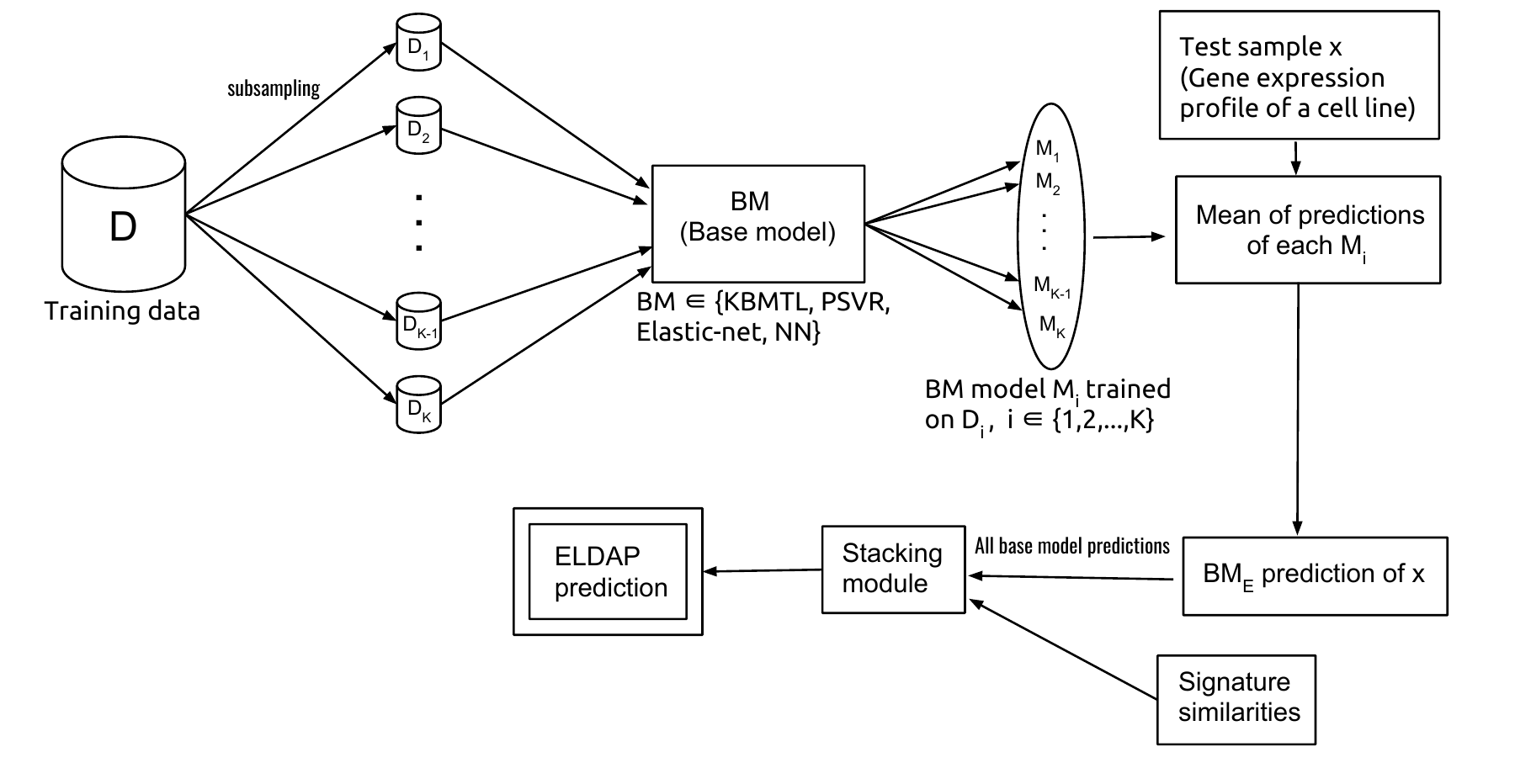}}
\caption{Block diagram for the ELDAP method\label{eldap-fig}. The prediction of ELDAP is the output
of a stacking module, where the input to the stacking module is the output of base model ensembles.
Some details are omitted in this figure, further details are given in the text.}
\end{figure}

Cancer types may substantially differ in aspects such as development, behavior and response to treatment. 
To consider this in the method, we also exploit some form of subsampling
by clustering the cell line samples. Our basic motivation here is to model the more closely related 
cancer types together, which can make the predictive models more stable and focused to certain types of 
cancer. Also, this operation can reduce the variance error of the final model.  To produce this final model 
by integrating the outputs of multiple learners, stacking technique with different meta-learners is used. Following 
subsections detail the proposed ensemble method. The whole method that we call Ensemble Learning for Drug 
Activity Prediction (ELDAP) is summarized in Figure \ref{eldap-fig}. We will refer the ELDAP method that 
does not exploit the signatures in stacking as ELDAP\textbackslash S. It has two different data sources for 
modeling cells. The training data (shown as D in Figure \ref{eldap-fig}) is the non-interventional gene expression 
profiles of cancer cell lines. On the other hand, interventional (drug-induced) gene expression data are exploited 
to compute the signature similarities.

\subsection{Base regressors}

This section defines the base models that we use as the learning methods in ensemble learning. These four 
are selected based on their previous use or their potential to be successful for this problem.

\noindent\textit{Elastic Net:} Elastic Net Regression which combines ridge and Lasso regression is a popular sparse 
regularized learning method. In genomics data sets,where number of features are typically larger 
than the number of samples, sparsity is a desired property. Elastic Net outputs a matrix $W$ that minimizes 
the following error function:

\begin{equation}
\frac{1}{2n}||Y - XW||^2_2
+  \alpha (\rho ||W||_1
+ (1 - \rho) ||W||^2_2)
\end{equation}

The $\alpha$ parameter sets the weight for the regularization. $\rho$ balances the lasso and ridge penalty and 
$n$ is the number of samples. These parameters were chosen as the ones minimizing the cross-validation error on a 
separate grid search. We used Elastic-net as a single task method where it is the model that was previously 
applied on GDSC for biomarker detection successfully \citep{garnett2012}.

\noindent\textit{Kernelized Bayesian Multitask Learning:}
Kernelized Bayesian Multitask Learning (KBMTL) \citep{Gonen2014} aims at solving related tasks jointly through a multitask regression scheme followed by a kernelized non-linear dimensionality reduction stage. The multitask algorithm reveals the shared information between distinct tasks in the kernel space and generates joint models. Such a procedure is capable of eliminating drug-specific noise as well as off-target effects. Moreover, joint modeling enables to utilize missing data since the regression is performed over a large set of data rather than individual data points. This property makes KBMTL a useful tool for genomics data where scarcity is a serious issue. KBMTL optimization procedure includes determining the hidden representations of data points in the kernel subspace and projection matrices.

\noindent\textit{Pairwise support vector regression:}
We call a support vector machine (SVM) a pairwise support vector regressor (PSVR) when the input data are composed of objects including two 
(possibly different) elements and the target is continuous. Depending on the task, a sample, therefore, usually represents an interaction or relationship 
between the elements. In our case, each sample is a drug-cell line pair and the target is the $IC_{50}$ value. 
This has been applied to a similar classification task before \citep{Tan2014DrugProfiles}. We follow here a similar procedure 
with the difference that our task here is a regression task. A similar approach 
was carried out also in protein-ligand interaction prediction \citep{Jacob2008ProteinApproach} and in compound toxicity prediction \citep{Bernard2017KernelToxicogenetics}.

We define a pairwise kernel $K_p$ as the product of the drug kernel $K_d$ and the 
cell line kernel $K_c$. 

\begin{equation}
K_p((dr_j, CL_i), (dr_l, CL_k)) = K_d(dr_j, dr_l) * K_c(CL_i, CL_k)
\end{equation}

\noindent where both $K_d$ and $K_c$ are defined as the radial basis function
(RBF) kernel between the drug and cell line descriptors, respectively. The descriptors 
that we use for drugs are the PaDEL descriptors \citep{yap2011}, and
the gene expression profile of the selected subset of genes for the cell lines.

\noindent\textit{Neural networks:} Artificial Neural Networks are one of the most widely used 
models in machine learning. 
With recent advancements in graphics processing units and training methods remarkable 
achievements have been witnessed in speech recognition, image understanding and machine 
translation. For an introduction and recent developments, see \citep{Goodfellow2016DeepLearning--book}.
In this paper, we used a multi-layer feed forward neural network with a multi-task output layer. 
The architectural details for the network that we use is given in Section \ref{experimental-sec}.

\subsection{Learning Strategy for base models}
\label{learn-strat}

We have followed a novel learning strategy with the intent of exploiting sample similarities. Instead of training a model or an ensemble of models trained on the whole training set, we generated several smaller training subsets corresponding to the neighborhood of a certain training sample. 

This training process can be summarized as follows: A single training sample is chosen and together with its 300 closest neighbors in the training set, we form a subset. 300 is chosen as the best performing value out of a set of neighborhood values. The distance metric defining the neighborhood of a sample is the Pearson correlation of the gene expression data. Due to the relatively small number of samples, with the target values this constitutes a highly under-determined system. To resolve this problem, we computed the Radial Basis Function (RBF) kernel gram matrix for all the methods except the PSVR. In this way, submatrix becomes a symmetrical square matrix where each index reflects the RBF similarity between two samples.  This not only performs a non-linear transformation, but also reduces the dimension. The models are trained on this newly formed gram matrix. We construct submatrices this way from the training set which, after training, results in a model corresponding to each sample. The testing of new samples is straightforward. When a test sample, which is a gene expression profile characterizing a cell line, arrives, the most similar 300 training cell lines and associated models are identified. The mean value of the predictions of this collection of models is returned as the overall predicted $IC_{50}$ value.

\subsection{Stacked generalization}

Stacked generalization (or stacking) method where individual predictions of a collection of models are introduced as inputs to a second-level learning algorithm is an extensively used technique for boosting prediction accuracy
\citep{Wolpert1992StackedGeneralization}. In our work, we blended Elastic Net, KBMTL, Neural Networks and PSVR through a stacking module. Although, stacking only aims at blending previously made predictions, addition of newer features to the process is also possible. As previously stated, we hypothesize that employing cell-line/drug signatures at the second level can augment the accuracy. For this purpose, we generated a new feature called signature similarity which is defined by the cosine similarity between drug and cell-line signatures. This feature is computed for all cell-line and drug pairs for which the signatures could be computed.


The main challenge of designing a second-level learner for blending was the missing values. Due to the amount of intersection between databases, we could acquire signature similarity metrics only for 133 drugs and 45 cell lines which approximately corresponds to 2.5 \% of the original data set. Therefore, we attempted to implement stacking schemes where signature similarity values contribute to the results only when they are available. This was a challenging task where the proposed models were supposed to be robust to changing input characteristics, e.g. missing/present signature similarity values. We propose three strategies, each satisfying these conditions. These can be splitted into three categories:

\noindent(i) \emph{Simple model averaging:} Here, the predictions of the individual models were averaged with and without weighting. When used, the weights were chosen as being inversely proportional to individual models’ MSEs on the data. In other words, a model making better predictions alone had more contribution in the average, as well. It is noteworthy that the signature similarities could not be directly included in the averaging since they did not represent IC$_{50}$ values but similarities. Therefore, prior to averaging, we utilized a pre-learner responsible for mapping signature similarities to IC$_{50}$ values. This pre-learner was chosen among a set of models including linear regression, support vector regression and regression tree. The choice and the tuning of this model were based on cross-validation. 

\noindent (ii) \emph{Conventional machine learning models:} Though variable input characteristics hampered the usage of classical machine learning models in a straight-forward way, it was still possible to train a different model for each input setting. For instance, one could train a regression tree model for observations with all possible features such as individual predictions and signature similarities, while turning to a new regression tree model for the samples lacking any of these features. For this class, we included models such as support vector machines, regression trees and shallow neural networks.

\noindent (iii) \emph{Modified distance-based models:} The last category of methods were the adaptive models which could dynamically adapt themselves to the changing input environments. In this class, we only included distance-based models since average inter-sample distances could be computed even in the absence of specific features. For this purpose, we designed a modified KNN regressor measuring the distances only based on the available features. 

For both data sets, we aimed to obtain the lowest MSE by varying these stacking strategies. It should be noted that, this process was fully automated, a meta-learner or combination of multiple meta-learners minimizing the cross-validation error was chosen from the bag of candidate models.  Same models were tested with and without signature similarity values so as to confirm the effect of similarities during prediction. For GDSC data set, the lowest MSE values were achieved as the average of two stacking modules; a modified KNN (iii) and model averaging (i). It should be noted that the signature similarities were transformed into IC$_{50}$ values by a linear model. In CCLE data set, an averaging module with a regression tree pre-learner was the most successful. No further averaging was found to be efficient. The effect of signature similarities was evaluated by comparing the MSE of indices, e.g. cell line-drug pairs, with available signature similarity values. We iterated the prediction process with and without signature similarities on these indices and observed the improvement when similarities were utilized.

\section{Experimental Results}
\label{experimental-sec}

In this section we report the experimental results corresponding to two different evaluations. In the first part, 
we report the results of the cross validation (CV) experiments that we name as \emph{in silico} experiments. The second part 
gives the results of \emph{in vitro} experiments of the selected cell line and drugs.

\subsection{\emph{In silico} evaluation}

For the \emph{in silico} evaluation, all the reported results are the average mean squared errors of a 3-fold CV. The experiments 
were performed on a computer with 3.2GHz CPU and 16GB of RAM running Linux. Here, for each compound, there is a vector of cytotoxicity values 
corresponding to the target vector of a learning problem. We refer to the drug activity prediction problem composed of a set of samples and 
one of these target vectors, as a task here. The performance of a task is measured in terms of the task specific mean squared error (TMSE);

\begin{equation}
TMSE(t) = \frac{1}{n} \sum_{i=1}^n (\hat{y}_i^t - y_i^t)^2
\label{tmse}
\end{equation}

\noindent for a given task $t$, where $n$ is the number of samples and $\hat{y}_i^t$ and $y_i^t$ are the predicted and 
correct cytotoxicity values for task $t$ and sample $i$, respectively. The overall performance is reported as the mean TMSE (MSE):

\begin{equation}
MSE = \frac{1}{T} \sum_{t=1}^T TMSE(t)
\label{mse}
\end{equation}

\noindent where $T$ is the number of tasks. In addition to MSE, we report the Spearman correlation of the predictions with the 
ground truth values. This is also computed similar to MSE, where the squared difference in Equation~\ref{tmse} is replaced by the 
Spearman correlation between $\hat{y}_i^t$ and $y_i^t$, $\rho(\hat{y}_i^t, y_i^t) $. Then, the overall correlation, $\overline{\rho}$, 
is computed as the mean correlation of all tasks as in Equation~\ref{mse}.

For each of the methods, we optimized the method parameters with a separate grid search. As a result of that, the parameters that 
we use are as follows. For Elastic-net, we used $\alpha=0.01$ and $\rho=1$ (which means 
that the method actually performs a Lasso regression). We used the scikit-learn \citep{Pedregosa2012Scikit-learn:Python} library for 
the implementation of Elastic-net. For KBMTL, we set $\{\alpha_\lambda, \beta_\lambda, \alpha_\epsilon, \beta_\epsilon, R, \sigma_h, \sigma_w\} = 
\{1,1,1,1,20,0.1,1\}$. We executed KBMTL for 200 
iterations. The author provided implementation was used for KBMTL. For neural network, we used two 
hidden layers with Rectified Linear Unit activations. 
For each hidden layer we used a dropout rate of 0.5. The size of the two hidden layers are 270 and 200 
respectively. We used the stochastic gradient descent optimizer with a learning rate of 0.01 and applied
a batch size of 100. We used Keras library \citep{Chollet2015Keras} for the implementation.


We report the results in five steps. In the first part, the performances of the single methods are given in Table
\ref{single-ensemble-a}. These results are reported as the baseline results for comparison. 
Due to the scalability issues of SVMs, the results for PSVR could not be reported for GDSC, as 
the run lasted for more than two weeks. For CCLE, as the data set is smaller, the performance of PSVR is also reported and 
it seems to be the best among others. However, one also should note that PSVR uses the extra information of compound structures.
Also, as mentioned before, the number of compounds with available structure data is 21 and 224 for CCLE and GDSC, 
respectively. One should take this into account for performance comparison. Among single methods, KBMTL significantly outperforms
other methods for GDSC. 

In the second step of the results we report in Table \ref{single-ensemble-b}, the performances of the single method ensembles 
(see Section \ref{learn-strat}) are given. We mark these methods with a subscript $E$ for readability as in Figure \ref{eldap-fig}. Except PSVR, 
for each method, we observe a performance improvement compared to Table \ref{single-ensemble-a}. Also note that, the improvement 
for Elastic-net is the largest. The performance difference is significant 
for Elastic-net and NN for GDSC. However it is not possible to confirm statistically that the improvements are significant for CCLE.

\begin{figure}[htbp]
\centering\includegraphics[scale=0.4]{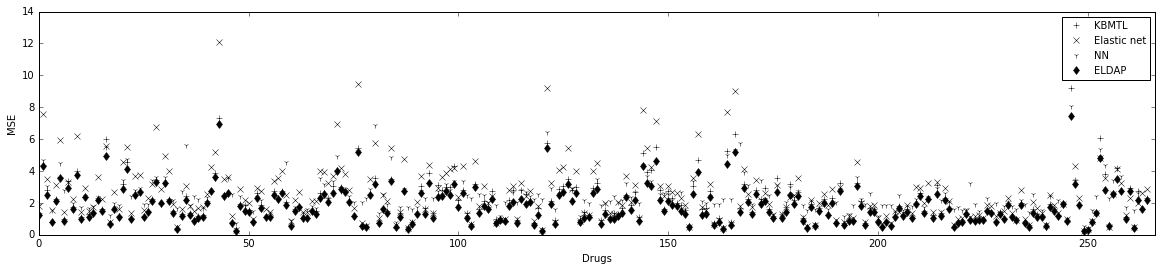}
\caption{Drug specific MSE for the GDSC data set. \label{task-spec-gdsc}}
\end{figure}

\begin{figure}[htbp]
\centering
\includegraphics[scale=0.4]{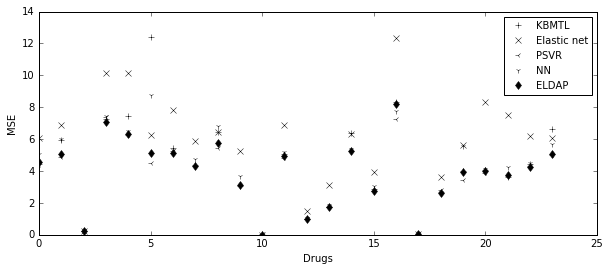}
\caption{Drug specific MSE for the CCLE data set. \label{task-spec-ccle}}
\end{figure}

The third part of the results report the overall performance of ELDAP and ELDAP\textbackslash S in Table \ref{eldap-res-a}. 
We can not see a performance improvement here, as the ratio of cell lines and drugs 
that have the signatures available is low. To demonstrate the effect of signatures, therefore, we report 
the results for the ones with signatures only.
In this case, a small improvement can be observed (see Table \ref{eldap-res-b}). We also provide some literature based evidence for some pairs 
with highly similar signatures in last part of this section. The fourth part of the results correspond to the drug specific errors. 
For each drug, we show the errors in Figure \ref{task-spec-gdsc} and \ref{task-spec-ccle} for GDSC and CCLE respectively. As it is seen in 
figures, the performance of ELDAP is superior in almost all drugs, especially for GDSC. Note that Figures \ref{task-spec-gdsc}
and \ref{task-spec-ccle} compare the single method performances (Table \ref{single-ensemble-a}) with ELDAP (Table 
\ref{single-ensemble-b}). 

Finally, Tables~\ref{single-ensemble-corr} and \ref{eldap-res-corr} report the results in terms of $\overline\rho$. In terms of 
$\overline\rho$, for GDSC, ELDAP overperforms the single methods (Table~\ref{eldap-res-a-corr}). For CCLE, the results are 
competitive with the baselines.

So far, we presented the efficiency and reliability of generated signature similarity metrics by
only inferring the augmented prediction accuracy when they are included in the stacking
stage. However, pharmalogical inferences are also possible. In this section, we attempt to
relate our results with the literature on specific drug activities. For this purpose, we randomly
select few cell line-drug pairs among the ones with very high/low signature similarity scores,
e.g. the best and the worst 5\% both for up and down regulations.
Among this short list, the most enhanced pattern is domination of few specific drugs on
several cell lines. In order words, the independent from the cell lines they are applied to, some
drugs achieved the highest signature similarity scores.
Majority of these succesful drugs were observed to target major pathways enhancing cancer
cell proliferation such as ERK/MAPK signaling and PI3K/MTOR signaling. Apart from
these, drugs regulating cell cycles and apoptosis through cycline depending kinases (CDK2,
CDK7, CDK9) and caspase activators (Procaspase-3, Procaspase-7), respectively, were also
present among the most successful drugs. The last set of drugs found among the top-tier drugs
were growth factor receptor inhibitors obstructing IGFR signaling (IGF1R, IR) and thus
cancer cell proliferation.
Obtaining high signature similarity results on these targeted drugs acting on the most
important pathways linked to cancer cell survival is biologically sensible and an indicator of
reasonability of our signature-based approach.

As a final evaluation, we report the performances of two "external" methods. These methods are SRMF and 
Random Forests (RF). We call them external to emphasize that they are not included in the ensemble for ELDAP. The results 
are given in Table~\ref{srmf-rf}. We evaluated these methods in the same setting that we used for the other methods. 
Before evaluating, we performed parameter optimization for both methods based on a separate validation set. 
Based on these results, we observed that RF overperforms other methods for both data sets. It is in our future plans to 
extend ELDAP with new methods and RF seems to be a good candidate to be included in the ensemble for ELDAP. 
For SRMF, we see that it is competitive for GDSC, but for CCLE we obtained poor results. However one should note that 
SRMF was designed originally to predict the activities for cells and drugs that were already 
available in the training set. Therefore SRMF is not suitable to predict drug activities for new cell lines which is 
the way we evaluate the methods in this paper. 

\begin{table}
\begin{subtable}{0.5\textwidth}
\begin{tabular}{p{2.1cm}p{2.1cm}p{2.1cm}}
\toprule
& GDSC & CCLE \\
\toprule
 KBMTL & 2.030$\pm$0.150 & 4.496$\pm$1.125 \\
 PSVR & - & 3.768$\pm$0.587 \\
 Elastic-net & 2.571$\pm$0.150  & 5.694$\pm$0.924 \\
 NN & 2.464$\pm$0.271 & 4.408$\pm$0.726 \\
\bottomrule
\end{tabular}
\caption{Single methods\label{single-ensemble-a}}
\end{subtable}
\begin{subtable}{0.5\textwidth}
\begin{tabular}{p{2.1cm}p{2.1cm}p{2.1cm}}
\toprule
& GDSC & CCLE \\
\toprule
KBMTL$_E$ & 2.064$\pm$0.150 & 4.073$\pm$0.522 \\
PSVR$_E$ & 2.518$\pm$0.205 & 4.806$\pm$1.040 \\
Elastic-net$_E$ & 2.015$\pm$0.171 & 4.233$\pm$0.736 \\
NN$_E$ & 2.040$\pm$0.156 & 4.087$\pm$0.695 \\
\bottomrule
\end{tabular}
\caption{Single method ensembles\label{single-ensemble-b}}
\end{subtable}
\caption{Performance of the single methods and single method ensembles in terms of MSE\label{single-ensemble}}
\end{table}

\begin{table}
\begin{subtable}{0.5\textwidth}
\begin{tabular}{p{2.1cm}p{2.1cm}p{2.1cm}}
\toprule
& GDSC & CCLE \\
\toprule
ELDAP\textbackslash S & 1.808$\pm$0.142 & 3.928$\pm$0.606 \\
ELDAP & 1.808$\pm$0.142 & 3.928$\pm$0.606 \\
\bottomrule
\end{tabular}
\caption{Overall performance\label{eldap-res-a}}
\end{subtable}
\begin{subtable}{0.5\textwidth}
\begin{tabular}{p{2.1cm}p{2.1cm}p{2.1cm}}
\toprule
& GDSC(Sig) & CCLE(Sig) \\
\toprule
ELDAP\textbackslash S & 1.880$\pm$0.553 & 1.840$\pm$0.544 \\
ELDAP & 1.971$\pm$0.773 & 1.816$\pm$0.687 \\
\bottomrule
\end{tabular}
\caption{Performance on signature sub-matrix\label{eldap-res-b}}
\end{subtable}
\caption{Performance of ELDAP in terms of MSE\label{eldap-res}}
\end{table}

\begin{table}
\begin{subtable}{0.5\textwidth}
\begin{tabular}{p{2.1cm}p{2.1cm}p{2.1cm}}
\toprule
& GDSC & CCLE \\
\toprule
 KBMTL & 0.159 & 0.244 \\
 PSVR & - & 0.212 \\
 Elastic-net & 0.159  & 0.231 \\
 NN & 0.170 & 0.235 \\
\bottomrule
\end{tabular}
\caption{Single methods\label{single-ensemble-a-corr}}
\end{subtable}
\begin{subtable}{0.5\textwidth}
\begin{tabular}{p{2.1cm}p{2.1cm}p{2.1cm}}
\toprule
& GDSC & CCLE \\
\toprule
KBMTL$_E$ & 0.159 & 0.234 \\
PSVR$_E$ & 0.155 & 0.191 \\
Elastic-net$_E$ & 0.159 & 0.205 \\
NN$_E$ & 0.170 & 0.224 \\
\bottomrule
\end{tabular}
\caption{Single method ensembles\label{single-ensemble-b-corr}}
\end{subtable}
\caption{Performance of the single methods and single method ensembles in terms of $\overline\rho$ \label{single-ensemble-corr}}
\end{table}

\begin{table}
\begin{subtable}{0.5\textwidth}
\begin{tabular}{p{2.1cm}p{2.1cm}p{2.1cm}}
\toprule
& GDSC & CCLE \\
\toprule
ELDAP\textbackslash S & 0.171 & 0.229 \\
ELDAP & 0.171 & 0.228 \\
\bottomrule
\end{tabular}
\caption{Overall performance\label{eldap-res-a-corr}}
\end{subtable}
\begin{subtable}{0.5\textwidth}
\begin{tabular}{p{2.1cm}p{2.1cm}p{2.1cm}}
\toprule
& GDSC(Sig) & CCLE(Sig) \\
\toprule
ELDAP\textbackslash S & 0.311 & 0.354 \\
ELDAP & 0.315 & 0.291 \\
\bottomrule
\end{tabular}
\caption{Performance on signature sub-matrix\label{eldap-res-b-corr}}
\end{subtable}
\caption{Performance of ELDAP in terms of $\overline\rho$ \label{eldap-res-corr}}
\end{table}

\begin{table}
\centering\begin{tabular}{p{2cm}p{3.5cm}p{3.5cm}}
\toprule
& GDSC & CCLE \\
\toprule
SRMF & 2.359$\pm$ 0.182 (0.160) & 9.318$\pm$2.173 (0.167) \\
RF & 1.709$\pm$0.137 (0.194) & 3.947$\pm$0.637 (0.225) \\
\bottomrule
\end{tabular}
\caption{Performance of SRMF and RF in terms of MSE ($\overline\rho$)\label{srmf-rf}}
\end{table}

\subsection{\emph{In vitro} evaluation}

In addition to the \emph{in silico} evaluation, we performed \emph{in vitro} evaluation 
on a cancer cell line. We selected A549, human epithelial lung carcinoma cell line, as our \emph{in vitro} test bed. Other than the 
availability of the cell line, this choice is arbitrary. 
Five different compounds were selected 
that were not applied on A549 in the GDSC data set. The selection was done by checking the error rates in the \emph{in silico} evaluation 
of the compounds and cytotoxicity variance of the drugs. The compounds are the ones that have the smallest error rates and largest variances 
(as variance can be considered an indicator of selectivity) among the commercially available chemicals. A549 was treated with these
five different compounds and the resulting cytotoxicity values were compared to the predictions by the proposed method. Following subsections 
give the results and further details of these wet-lab experiments.

\subsubsection{Chemicals}
The chemicals and reagents used in the experiments were purchased from the following suppliers: Fetal calf serum (FCS), trypsin–EDTA, Dulbecco's modified Eagle's medium (DMEM), penicillin-streptomycin, L-glutamine from Biological Industries (Kibbutz Beit-Haemek, Israel); 3-(4,5-dimethylthiazol-2-yl)-2,5-diphenyltetrazolium bromide (MTT) dye, methylene blue dye, methanol (MEOH), from Merck Chemicals (Darmstadt, Germany); dimethyl sulfoxide (DMSO), ethanol, from ICN Biomedicals Inc. (Aurora, Ohio, USA). 

\subsubsection{Cell culture}

A549 cells, human epithelial lung carcinoma cell line, were purchased from American Type Culture Collection (Manassas, VA, USA). The cells were maintained in DMEM containing 10\% fetal calf serum and 0,5\% penicillin-streptomycin. The cells were kept in a incubator (Thermo Scientific Heraeus, Germany) at 37 $\pm$ \ang{1}C with a humidified atmosphere of 95\% air and 5\% CO$_2$. The culture medium was changed two times in one week. A549 cells used in all experiments were between the 2nd and 3rd culture passage after thawing.

\subsubsection{Determination of cytotoxicity by MTT assay}
\label{in-vitro-method}

The effects of the compounds, TAK-715, GSK2126458, Ispinesib mesylate, KIN001-102 and OSI-930 on cell viability were determined by MTT assay with slight changes described by \citep{Ohguro1999ConcentrationCells}.  A549 cell line is highly sensitive and has good characteristics in colony formation. Therefore these adherent cells are very convenient for MTT assay. Before starting experiments, the cells have been grown for 2 weeks. Afterwards  A549 cells were seeded at density of 5000 and 10000 cells/well in 96-well plates and allowed to grow for 24 h before treatment. The cells were treated with each compounds at different concentrations (0.0064, 0.032, 0.16, 0.8, 4, 20, 100, 500 $\mu$M for TAK-715, KIN001-102, OSI-930 and  0.00128, 0.0064, 0.032, 0.16, 0.8, 4, 20, 100 $\mu$M for GSK2126458, Ispinesib mesylate) in the culture medium for 48 h and 72 h. The compounds were dissolved in DMSO and added to the medium to yield a final DMSO concentration of 1\% (v/v). Control group cells were treated with the culture medium containing 1\% of DMSO without compounds. Following the incubation period, the medium was removed and 1 mg/mL MTT solution in 100 $\mu$L of the culture medium was added to each well and further incubated at \ang{37}C for 3 h. After the MTT application, the medium was discarded, and 100 $\mu$L of PBS was added to wash the cells. 100 $\mu$L of DMSO was added onto the cells in order to dissolve the formazan crystals. Absorbance of each sample was measured at 570 nm using the microplate reader (SpectraMax M2, Molecular Devices Limited, Berkshire, UK). Cytotoxic potantial of the compounds were determined by calculating the percentage of cell viability ratio between treated and untreated (control) cells (\% cell viability) IC50 values of the drugs was calculated by using the dose-response curves. Each experiment was performed in triplicate.

\subsubsection{Evaluation of cytotoxic effects of the compounds on A549 cell line}

\begin{figure}[htbp]
\includegraphics{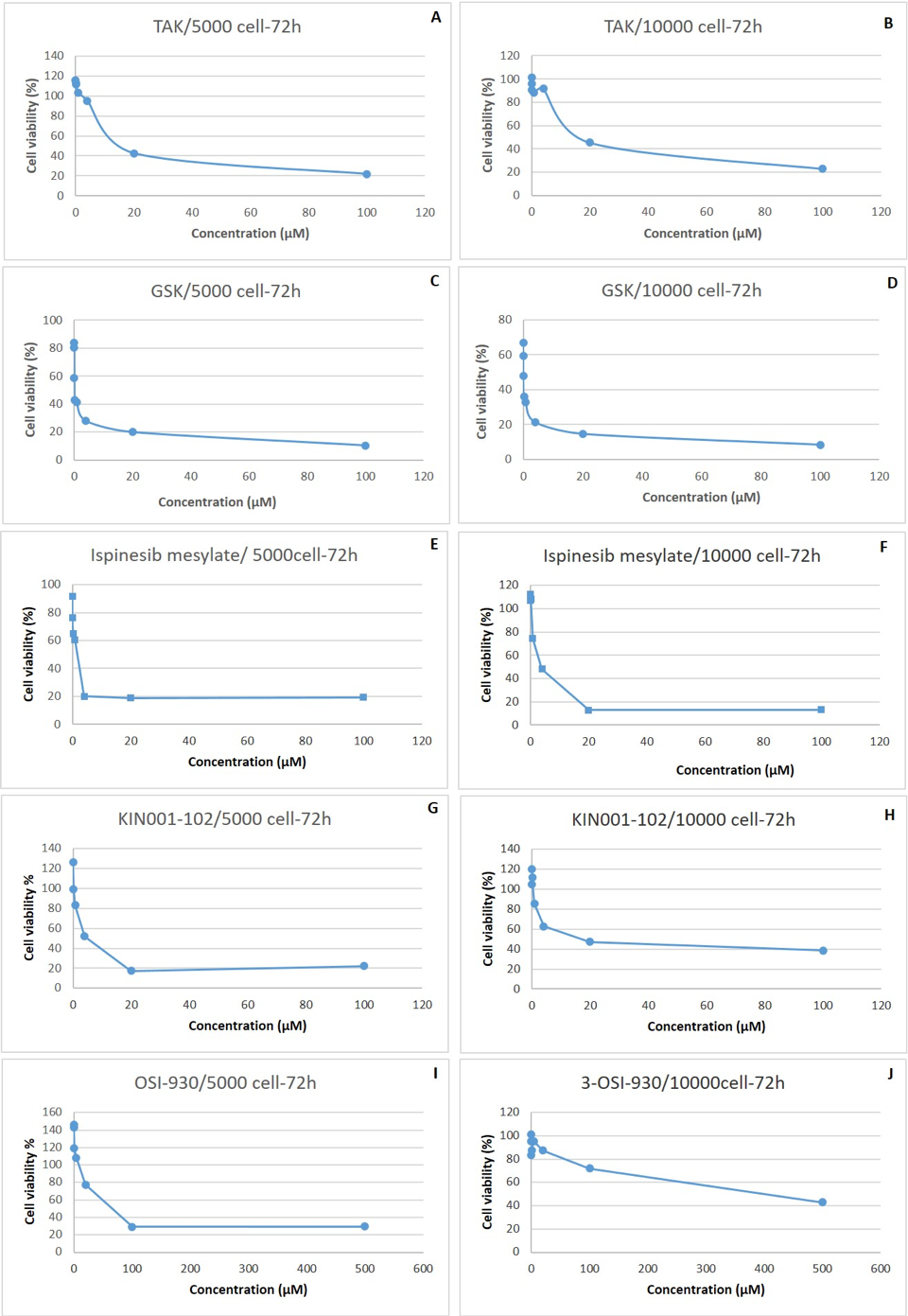}
\caption{Effects of the compounds on cell viability of A549 cells with 5000 cell/well (A, C, E, G, I) or 10000 cell/well (B, D, F, H, J) for 48 h exposure.\label{invitro-fig1}}
\end{figure}

\begin{figure}[htbp]
\includegraphics{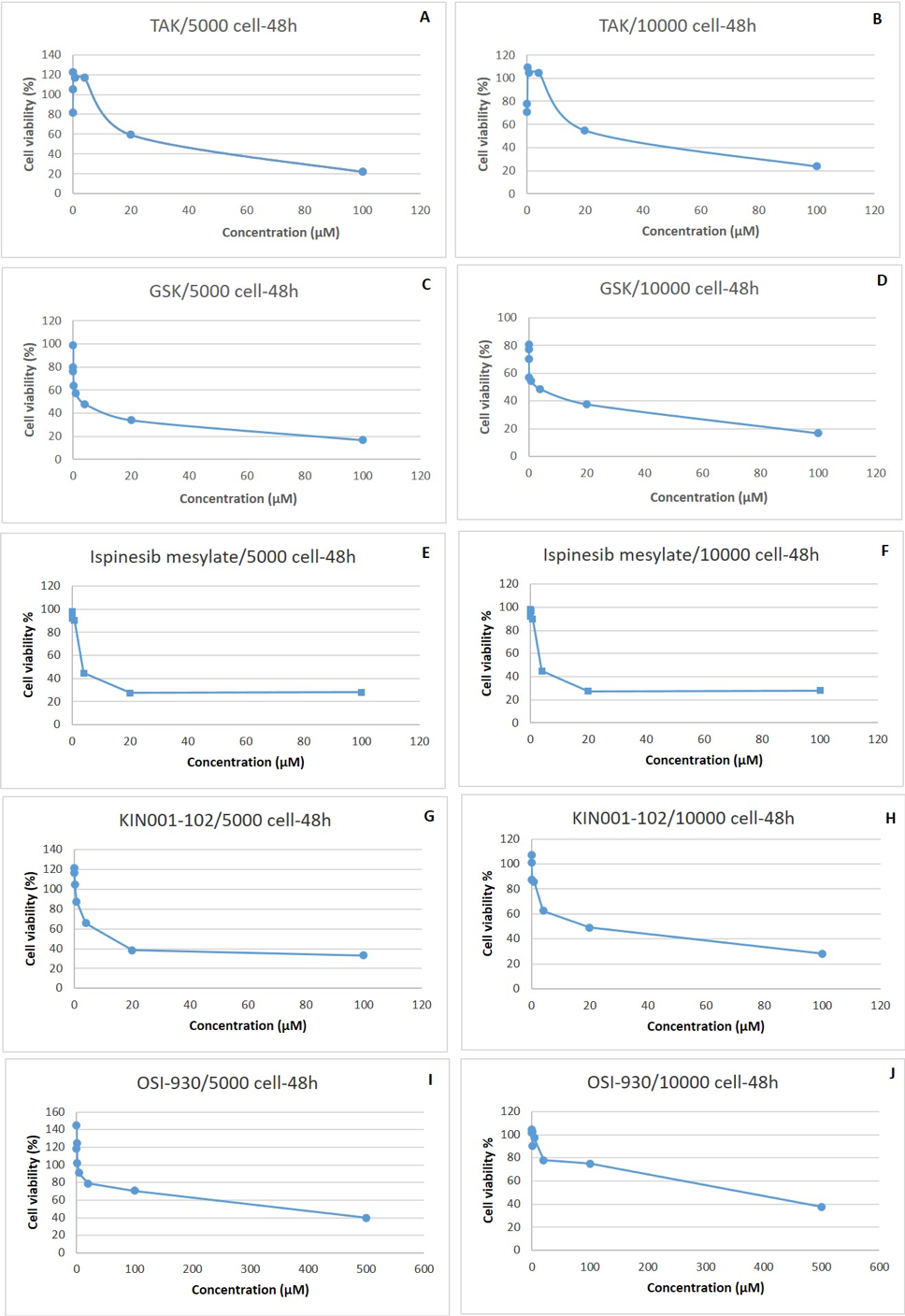}
\caption{Effects of the compounds on cell viability of A549 cells with 5000 cell/well (A, C, E, G, I) or 10000 cell/well (B, D, F, H, J) for 72 h exposure.\label{invitro-fig2}}
\end{figure}

The cytotoxic effects of the compounds at different concentrations on A549 cell line are shown in Figure \ref{invitro-fig1} and \ref{invitro-fig2}. According to the results, the compounds showed significant cytotoxicity on A549 cell line. The concentration-dependent cytotoxicity was observed for A549 cells after both 48 h and 72 h exposure to the compounds.

The IC$_{50}$ values of cytotoxic drugs were calculated as explained in Section \ref{in-vitro-method} and shown in Table 
\ref{invitro-tab1} and \ref{invitro-tab2}. The \emph{in vitro} results were compared to the predictions of ELDAP for 
all compounds. The predictions of ELDAP for these drugs are shown in Table \ref{predictions-tab}. As it is shown in Table \ref{invitro-tab1} 
and \ref{invitro-tab2} the IC$_{50}$ values of the compounds are inline with IC$_{50}$ predictions in Table 
\ref{predictions-tab}. Note that one should not expect an exact match with the predictions here as there are significant differences in experimental conditions though 
we did our best to build a similar wet-lab environment that the GDSC data set is produced.

\begin{table}[h]
\centering
\begin{tabular}{p{4cm}p{2.5cm}p{2.5cm}}
\toprule
\textbf{Drug} & \textbf{IC$_{50}$ (48 h)} & \textbf{IC$_{50}$ (72 h)} \\
\toprule
TAK 715 & 40.53 & 3.32 \\
\midrule
GSK 2126458 & 17.74 & 0.10 \\
\midrule
Ispinesib Meylate & 3.63 & 1.63 \\
\midrule
KIN001-102 & 13.26 & 5.01 \\
\midrule
OSI-930 & 370.31 & 65.34 \\
\bottomrule
\end{tabular}
\caption{The IC$_{50}$ ($\mu$M) values of the compounds for A549 cells with 5000 cell/well at 48h and 72 h\label{invitro-tab1}}
\end{table}

\begin{table}[h]
\centering
\begin{tabular}{p{4cm}p{2.5cm}p{2.5cm}}
\toprule
\textbf{Drug} & \textbf{IC$_{50}$ (48 h)} & \textbf{IC$_{50}$ (72 h)} \\
\toprule
TAK 715 & 32.39 & 18.40 \\
\midrule
GSK 2126458 & 3.22 & 0.03 \\
\midrule
Ispinesib Meylate & 3.63 & 3.77 \\
\midrule
KIN001-102 & 19.02 & 17.05 \\
\midrule
OSI-930 & 368.56 & 403.67 \\
\bottomrule
\end{tabular}
\caption{The IC$_{50}$ ($\mu$M) values of the compounds for A549 cells with 10000 cell/well at 48h and 72 h\label{invitro-tab2}}
\end{table}

\begin{table}[h]
\centering
\begin{tabular}{p{4cm}p{5cm}}
\toprule
\textbf{Drug} & \textbf{Predicted cytotoxicity} \\
\toprule
TAK 715 & 106.738 \\
\midrule
GSK 2126458 & 0.071 \\
\midrule
Ispinesib Meylate & 0.285 \\
\midrule
KIN001-102 & 21.716 \\
\midrule
OSI-930 & 69.036 \\
\bottomrule
\end{tabular}
\caption{Predicted cytotoxicity values by ELDAP\label{predictions-tab}}
\end{table}

\section{Conclusion}
\label{conc-sec}

Prediction of chemotherapeutic response of malign cells is an essential part of designing 
personalized drugs. Recent efforts of producing large scale data sets provides the 
opportunity to apply machine learning models to this problem. 

In this paper we proposed an ensemble learning method which combines both multi-task and 
single task learners through a stacking module. We demonstrated the usage of novel 
signatures that represent the activity and sensitivity of drugs and cell lines, respectively. 
These signatures are defined in a way that the similarity between them can be used as 
an indicator of drug activity for the corresponding cell line. In addition to the 
computer based evaluation, \emph{in vitro} experiments 
performed on cell line A549 for 5 drugs depicted that the method can produce reliable 
predictions. The small number of drugs in the intersection of LINCS and the other databases, 
GDSC and CCLE, limited the efficacy of the signature similarities. In case of the data 
availability, the significance of the effect can be more clearly demonstrated. 

Several directions can be followed to extend this work. One is to incorporate new data types 
that can effect the results, such as the pathway data sets. The pathway information can help in 
selecting the genes in the same pathway for modeling. Another usage can be to exploit the 
pathway cancer relationship in the sense that only related genes can be used in the sub-models. 
This can be further extended in case the cell lines are grouped by their tissue or cancer type.
Another possible extension is to apply active learning techniques to this problem, where 
the next wet-lab experiment to perform can be selected in such a way that the new 
cytotoxicity value is maximally effective on improving the prediction performance. We believe that 
this kind of a cycle can greatly decrease the error rates.

\bibliographystyle{plainnat}
\bibliography{Mendeley}

\end{document}